\documentclass[a4paper,nocompress]{spie}  
\usepackage[]{graphicx}

\title{3D-integrated beam combiner for optical spectro-interferometry} 

\author{Stefano Minardi\supit{a}, Allar Saviauk\supit{b}, Felix Dreisow\supit{a}, Stefan Nolte\supit{a}, and Thomas Pertsch\supit{a} 
\skiplinehalf
\supit{a}Institute of Applied Physics, Abbe Center of Photonics, Friedrich-Schiller-Universit\"at Jena, Max-Wien-Platz 1, 07743 Jena, Germany; \\
\supit{b} Leibniz-Institut f\"ur Astrophysik Potsdam, An der Sternwarte 16, 14482 Potsdam, Germany\\
}

\authorinfo{Further author information: S.M.: E-mail: stefano@stefanominardi.eu, Telephone: +49 (0)3641 947 848}

\begin{document} 
  \maketitle 

\begin{abstract}
We present a compact setup based on a three-dimensional integrated optical component, allowing the measurement of spectrally resolved complex-visibilities for three channels of polychromatic light.  We have tested a prototype of the component in R band and showed that accurate complex visibilities could be retrieved over a bandwidth of 50 nm centered at 650 nm (resolution: R=130). Closure phase stability in the order of $\lambda/60$ was achieved implying that the device could be used for spectro-interferometry imaging.
\end{abstract}


\keywords{Interferometry, beam-combiners, integrated optics}

\section{INTRODUCTION}
\label{sec:intro}  
Spectro-interferometry is an advanced astronomical technique used to deliver high-resolution images of astronomical targets and to study their morphological features. The technique consists in dispersing the light after a beam combiner and analyzing the complex visibilities as a function of the wavelength. 
Variations of the visibilities across the spectrum of an astronomical target can be exploited to devise its astrophysical morphology \cite{DdS:2007}. On the other hand, for more spectrally uniform objects, it is possible to exploit the dependence of the (u,v) coordinates on the wavelength to achieve a better Fourier-plane coverage for interferometric imaging applications \cite{Lacour:2008}.

In this context, it was recently shown that a 4-telescope integrated optical beam combiners \cite{Benisty:2009} 
coupled with an imaging spectrometer
are able to obtain high precision measurements of the visibility across the spectrum leading to very good (u,v)-plane coverage in just a few hours of observation \cite{Blind:2011}. However, to increase the imaging capabilities of interferometers, simultaneous combination of a larger number of 
telescopes is desirable. 
Unfortunately, conventional integrated optics components are not easily scalable to the simultaneous combination of more than 4-6 telescopes, since the planar constraint makes the design of the beam combiner increasingly complex.

This complexity inherent to the planar geometry could be lifted by arranging the waveguides in three dimensions. 
For instance, Rodenas et al. demonstrated in the laboratory the operation of a simplified three-channel, mid-infrared beam combiner featuring three-dimensional (3D) curved waveguides and Y-junctions written in chalcogenide glasses \cite{Rodenas:2012}. 
The first  interferometric 3D component to be tested on-sky was the Dragonfly photonic chip, where
waveguides are used to remap 4 pupil apertures into a non redundant linear array
of point-like sources for free-space multi-axial beam combination. The potential
of pupil remapping techniques resides in the possibility to retrieve high-dynamical
range images from seeing limited instruments and with a resolution close to the
diffraction limit \cite{Perrin:2006}. Finally,  Minardi et al. showed in laboratory
experiments \cite{Minardi:2012b} that the design of a beam combiner can be radically simplified 
down to a two-dimensional regular array of coupled waveguides (the discrete
beam combiner, DBC) \cite{Minardi:2010}. 

In this paper, we focus on the polychromatic operation of the DBC.
While the monochromatic operation of the device\cite{Minardi:2012b} is (in a sense) trivial, the strong wavelength dependence of waveguide coupling can affect the accuracy of phase and visibility retrieval when used with polychromatic light. 
We found that by operating the DBC in 'white light mode', the coherence retrieval works with bandwidths as large as 17 nm at a central wavelength of 640 nm \cite{Saviauk:2013}. This is obviously a too small bandwidth to be useful for astronomical applications. 
A solution to this problem is to use the DBC in spectro-interferometry mode and apply the coherence retrieval method to small bandwidths.  By dispersing the output of the array we could extend the operation of the DBC to a total bandwidth of 50 nm centered at 650 nm. The range is limited only by the output pitch of the array of waveguides which can be changed  by design. The capability of the component to retrieve visibility and phase is uniform across the whole investigated spectral range. In particular, the closure phase (sum of all possible phase differences between channels) was stable within $\lambda/36$, implying that high precision interferometric imaging is possible. 
Scalability to infrared wavelengths of the DBC component will be discussed as well as possible implementations for multi-telescope interferometric instruments.

\section{THEORETICAL BACKGROUND}

\subsection{Operation principle of the discrete beam combiner}
The discrete beam combiner exploits the properties of discrete diffraction \cite{Christodoulides:2003} to achieve interferometric beam combination with two dimensional arrays of evanescently coupled, single mode waveguides \cite{Minardi:2010}.
Because of the evanescent coupling, light injected in a single waveguide will excite neighboring waveguides of the array in a fashion resembling free-space diffraction, but bounded to a discrete number of sites (the waveguides) - hence the name 'discrete diffraction'. 
Each waveguide of the array will carry a fraction of the input flux of light, with a field amplitude and phase depending only on the position of the input waveguide and the length of the array. 
By exciting simultaneously more waveguides, the propagating fields due to each input waveguides will sum up coherently in the waveguides of the array giving raise to a discretized interference pattern. 

By arranging  in a vector $\vec{I}$ the output intensities at each waveguide (the output excitation pattern of the waveguide array) we can use a linear transformation (Visibility To Pixel Matrix or V2PM \cite{Tatulli:2007}, hereafter indicated with $\{\alpha\}$) to obtain the vector $\vec{J}$ containing the quadratures of the complex visibilities pertaining to all possible pairwise combinations of the input fields. Thus, by applying the (pseudo)-inverse of the V2PM (Pixel To Visibility Matrix, P2VM) to the readout of the discrete intensity pattern at the output of the waveguide array, the visibilities can be extracted:
\begin{equation}
\vec{J}=\{\alpha\}^{-1}\vec{I}
\end{equation}
A minimal condition for the (pseudo)-inverse of the  $\{\alpha\}$ matrix to exist is that the array should feature at least $N\times N$ waveguides, $N$ being the number of combined fields (corresponding to apertures/telescopes)\cite{Minardi:2012}. 
Even in that case, however, not all the choices of input sites and lengths of the waveguide array are suitable for interferometric beam combination. Indeed, most of the input configurations/array lengths result in badly conditioned or even singular V2PM. Thus it is necessary to identify (\textit{e.g.} by numerical search) the combination of input configuration and array length which leads to the most stable V2PM. 
In the numerical parametric scan, a useful quantity to gauge the stability of the $\{\alpha\}$ matrix is its condition number, defined as the ratio of the maximum to the minimum singular values \cite{Press}.  The smaller the condition number of the V2PM, the better is the stability of the DBC method against errors in the measurement of $\vec{I}$ and the calibration of the elements of the matrix $\{\alpha\}$.

\begin{figure}[hb]
   \begin{center}
   \begin{tabular}{c}
   \includegraphics[height=6cm]{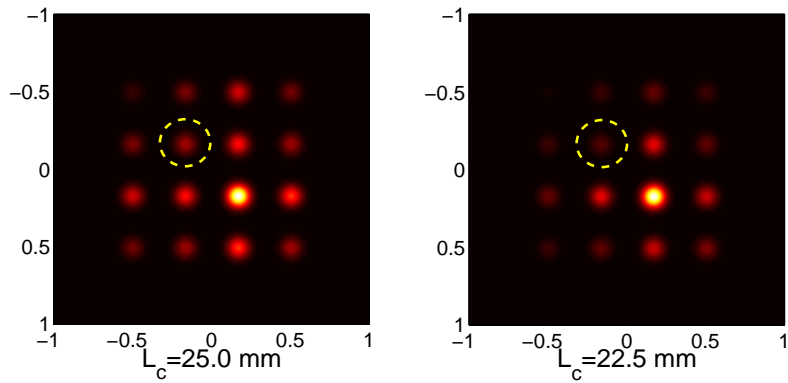}
   \end{tabular}
   \end{center}
   \caption[Integrated optic] 
   { \label{fig:coupling} 
Simulated effect of chromatic dispersion of the coupling strength in a the waveguide array of length $L=35$ mm. Left: discrete diffraction pattern for short wavelengths (coupling length $L_c=25$ mm), Right: discrete diffraction pattern for the same sample but at longer wavelength (coupling length $L_c=22.5$ mm). In both pictures the injection waveguide is highlighted by the circle.}
   \end{figure} 

\subsection{Chromaticity of the evanescent coupled waveguides}
As mentioned before, evanescent coupling of waveguides is strongly chromatic. The chromaticity arises from the fact that the size of the fundamental mode of the waveguide
is proportional to the wavelength \cite{SnyderLove}. As a consequence, the overlap integral between two modes of neighboring waveguides (which regulates their coupling strength) grows steeply with the wavelength. 
The variation of the coupling strength leads to a variation of the discrete diffraction pattern generated by each input waveguide (see Fig. \ref{fig:coupling}) which in turn modifies the elements of the V2PM.
Thus, for each wavelength $\lambda$ we can define
a corresponding matrix $\{\alpha\}_{\lambda}$ and a vector $J(\lambda)$. This
means that the same coherence properties of light will generate a different output pattern for
each wavelength. The overall output intensity at
each waveguide can be expressed as the integral over
the wavelength of the monochromatic intensities:
\begin{equation}
\label{eq:polychrom}
\vec{I}=\int_{\lambda_{\mathrm{min}}}^{\lambda_{\mathrm{max}}}\alpha_\lambda\vec{J}_\lambda d\lambda
\end{equation}
In this case, we cannot define the measured intensity
pattern in terms of a matrix by vector product. A
solution to come around this issue is to disperse the
output light so that we can apply the DBC method to
every single wavelength.
An alternative to solve Eq. (\ref{eq:polychrom}) is to have an
achromatic matrix $\{\alpha\}$, which is independent of wavelength.
In this case, the matrix can be factorized out
from the integral and we can retrieve the wavelength
averaged $J$ vector. In general, we can assume that
there is a certain bandwidth $\Delta\lambda$ such that the matrix
$\{\alpha\}$ behaves achromatically within a very good
approximation so that:
\begin{equation}
\vec{I}=\alpha\int_{\lambda_{\mathrm{c}}-\Delta\lambda/2}^{\lambda_{\mathrm{c}}+\Delta\lambda/2}\vec{J}_{\lambda}d\lambda=\alpha\vec{J}
\end{equation}
The wavelength averaged $J_{\lambda}$ vectors correspond to
the $J$ vector of a polychromatic source, since the
quadratures of the coherence functions for each
wavelength can add up.

\section{SETUP}

As we have seen in the previous section, a way to operate a DBC despite the chromatic dispersion of the evanescent coupling is to 
select a narrow observing bandwidth of the discrete interference pattern. Broadband operation is then ensured if we are able to scan the 
central wavelength of observation. Simultaneous acquisition of the interference patterns at several wavelengths is possible by exploiting a method already used to acquire data cubes for integral field spectroscopy \cite{AllingtonSmith:2006}.

By dispersing the light across the area in between the waveguides, it is possible to acquire the discrete interference pattern simultaneously for different wavelengths, the same way as lenslet-array-based integral field units do to acquire spectra of contiguous sampling points\cite{Bacon:2001}.
Because the evanescent coupling used for interferometric beam combination requires that the waveguide separation is only a few times the mode size, it is necessary to reformat the array of waveguides into an array of much larger pitch to accommodate the spectra.  
By analyzing the DBC pattern of light at each individual
color, it is possible to determine, wavelength by wavelength, the coherence properties of the combined light.

\begin{figure}[t]
   \begin{center}
   \begin{tabular}{c}
   \includegraphics[height=8cm]{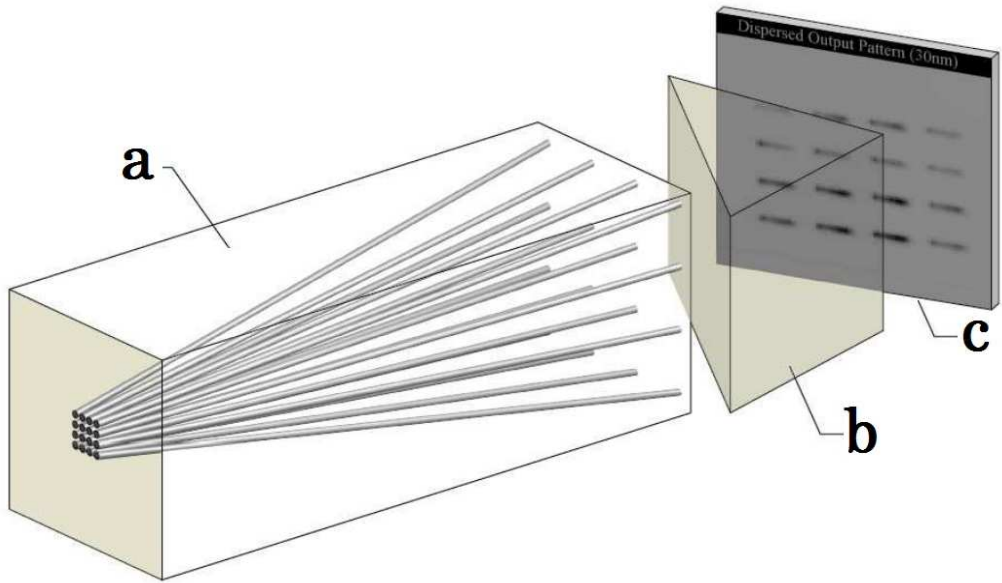}
   \end{tabular}
   \end{center}
   \caption[Integrated optic] 
   { \label{fig:beam_comb} 
Conceptual scheme of a DBC used for spectro-interferometry. a) A 4x4 DBC with  fan out waveguides. b) Dispersive stage (imaging spectrograph). c) Plane of the camera where the 4x4 spectra are projected.}
   \end{figure}

A conceptual scheme of the DBC component and its integration in a setup suitable for spectro-interferometry is illustrated in Fig. \ref{fig:beam_comb}.  At the beginning of the sample, a small pitch between the waveguides allows a strong inter-waveguide coupling
which permits the DBC operation. The coupling decreases as the waveguides are driven
apart, so that the discrete interference pattern formed at the beginning is frozen in the
waveguides. At the end of the sample the separation between waveguides is large
enough so that it is possible, by means of an imaging spectrograph (represented by a prism), to project low
resolution frequency spectra of the light coupled in each individual waveguide in
the gaps between them.

 \begin{figure}[b]
   \begin{center}
   \begin{tabular}{c}
   \includegraphics[height=8cm]{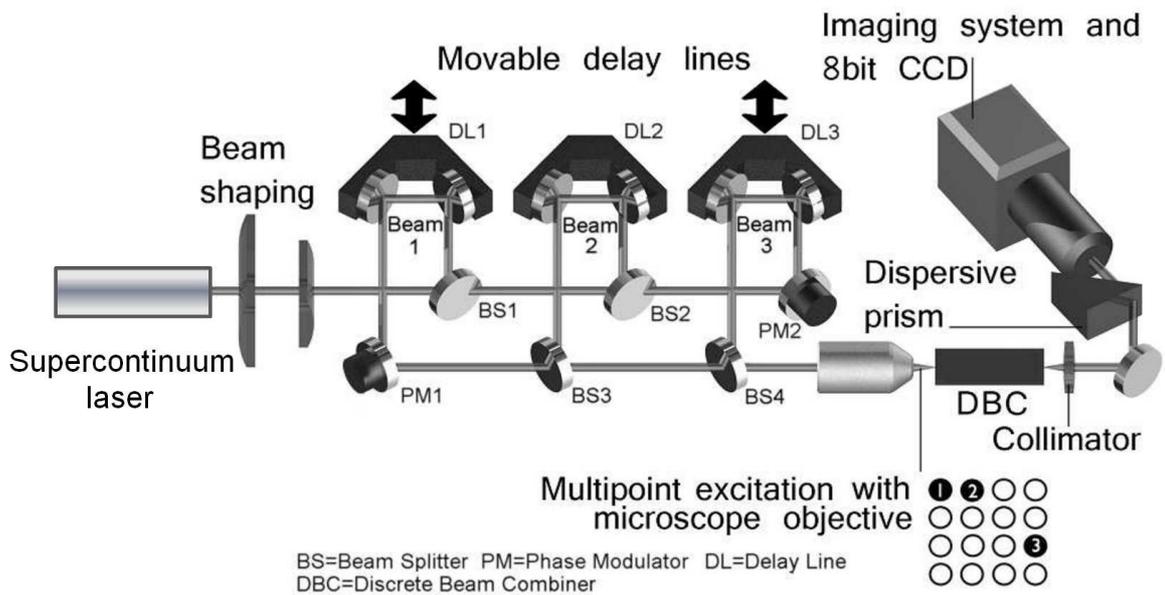}
   \end{tabular}
   \end{center}
   \caption[Integrated optic] 
   { \label{fig:setup} 
Implemented setup for the combination of 3 beams of polychromatic light by means of a 4x4 DBC component. BS: Beam Splitter; DL: Delay Line; PM: Phase Modulator.}
   \end{figure} 

We present now the laboratory characterization of a DBC prototype manufactured 
to combine simultaneously three telescopes and operate from  640 nm to 690 nm. This wavelength range was dictated by the
available polychromatic light source and optical equipment of our laboratory, but the results of the experiments are 
representative of what could be achieved at other wavelengths, for instance in the
near- or mid-infrared. The setup used to test the DBC is illustrated in Figure 3 and is composed by four main units: 1) a polychromatic light source, 2) a three channel Mach-Zehnder interferometer, 3) the DBC component and 4) a imaging spectrometer.

The polychromatic light source is a supercontinuum laser from which up to 8 channels of bandwidth $\Delta\lambda=2.5 $nm can be independently selected within a range spanning from 640 to 800 nm. The output is coupled to a single mode fiber which acts as a spatial filter and is fed to the setup.

After collimation, the beam from the polychromatic light source is shaped by a pair of cylindrical lenses in order to match the elongated beam shape of the waveguide modes.

The beams of the three telescopes were simulated by the three channels of the
modified Mach-Zehnder interferometer of Fig. \ref{fig:setup}. The complex visibilities arising
from the interference of pairs of beams chosen from the three channels represent
the visibilities of a baseline (baseline1=Beam1-Beam2, baseline2=Beam1-Beam3,
baseline3=Beam2-Beam3). The three beams are focused on the input waveguides
of the DBC component, as illustrated in the inset picture. 
The phase between the beams could be controlled by means of phase modulators (PM in Fig. \ref{fig:setup}) nested in two of the delay lines.

The integrated optical component consists of a tapered square 4x4 array of coupled waveguides. The pitch of the array varies linearly from 13 to 80 microns over a length of 75 mm. The output array is then dispersed by a miniature, prism-based imaging spectrograph with resolution R=130 and recorded by a CCD camera.  We adjusted the magnification of the imaging spectrometer so that the mode size of a single waveguide was matched to the pixel size of the CCD.

\section{FABRICATION OF THE DBC SAMPLE}
The technological platform for 3D photonics is based on the direct writing of
transparent materials with tightly focused ultrashort ($< 100$ fs) laser pulses. 
Under irradiation of high intensity laser pulses a localized, low density plasma is formed which develops, upon recombination,
in defects or a reconfiguration of the local material structure. 
These structural modifications manifest themselves as local variations of
the refractive index, which can be positive or negative depending on the type of
irradiated material. By scanning the laser beam focus inside the material with
a 3-axes positioning system, it is possible to inscribe complex refractive index
structures in 3D.
As sample material we chose fused silica, a material which exhibits positive refractive index variation upon irradiation and can 
thus be used to create waveguides with relatively low loss (about 0.5 dB/cm) in the visible region \cite{Felix}.

\section{RESULTS}

\begin{figure}
   \begin{center}
   \begin{tabular}{c}
   \includegraphics[width=12cm]{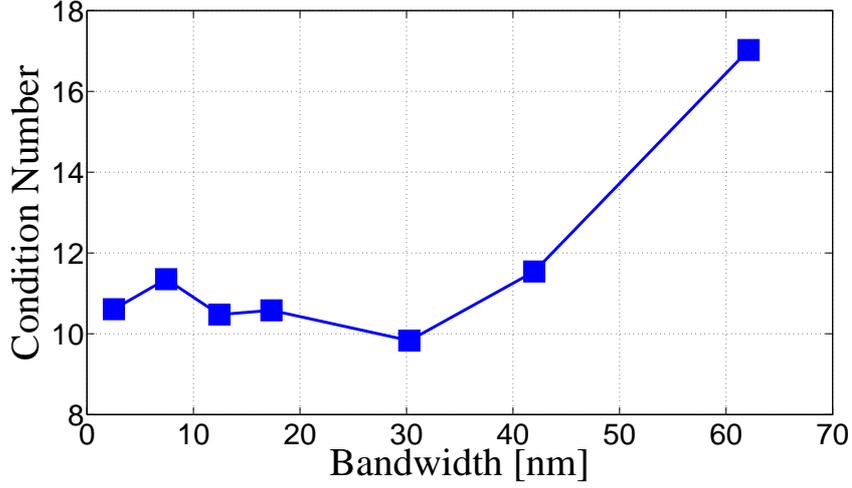}\quad
   \end{tabular}
   \end{center}
   \caption[Integrated optic] 
   { \label{fig:conditionnumber} Condition number of the calibrated V2PM of the DBC as a function of the bandwidth of the light. }
   \end{figure}

Our experimental characterization of the DBC component with polychromatic light 
started by analyzing the stability of the calibrated V2PM matrix (for the calibration procedure see Saviauk et al. 2013
\cite{Saviauk:2013}) for different bandwidths of the polychromatic light. The bandwidth of the 
light source was changed from a minimum of 2.5 nm to a maximum of 62 nm by selecting several channels 
of our supercontinuum laser. The stability of the V2PM was evaluated by calculating its condition number. 
The results are illustrated in Fig. \ref{fig:conditionnumber}. The condition number remains about constant ($\sim11$)
for bandwidths smaller than 40 nm, then it sharply increases. The condition number alone however does not 
account for the accuracy of the measurement of the coherence parameters, which was carried out with additional tests 
described in detail in Saviauk et al. 2013\cite{Saviauk:2013}. These additional tests pointed out that the maximum polychromatic bandwidth which can be used without dispersing the output of the DBC is about 17 nm, for a central wavelength of 640 nm.

\begin{figure}
   \begin{center}
   \begin{tabular}{c}
   \includegraphics[height=6cm]{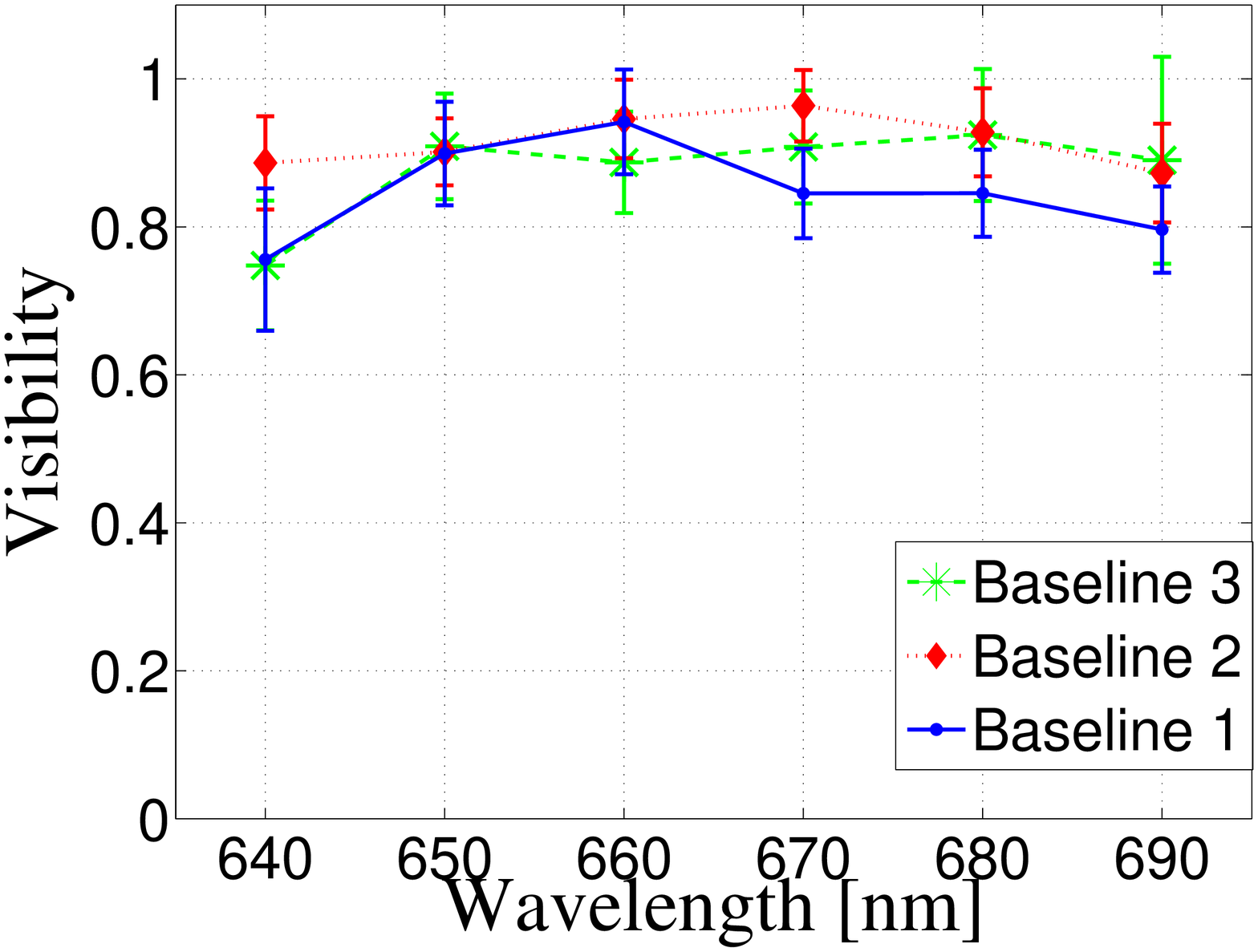}\quad
   \includegraphics[height=6cm]{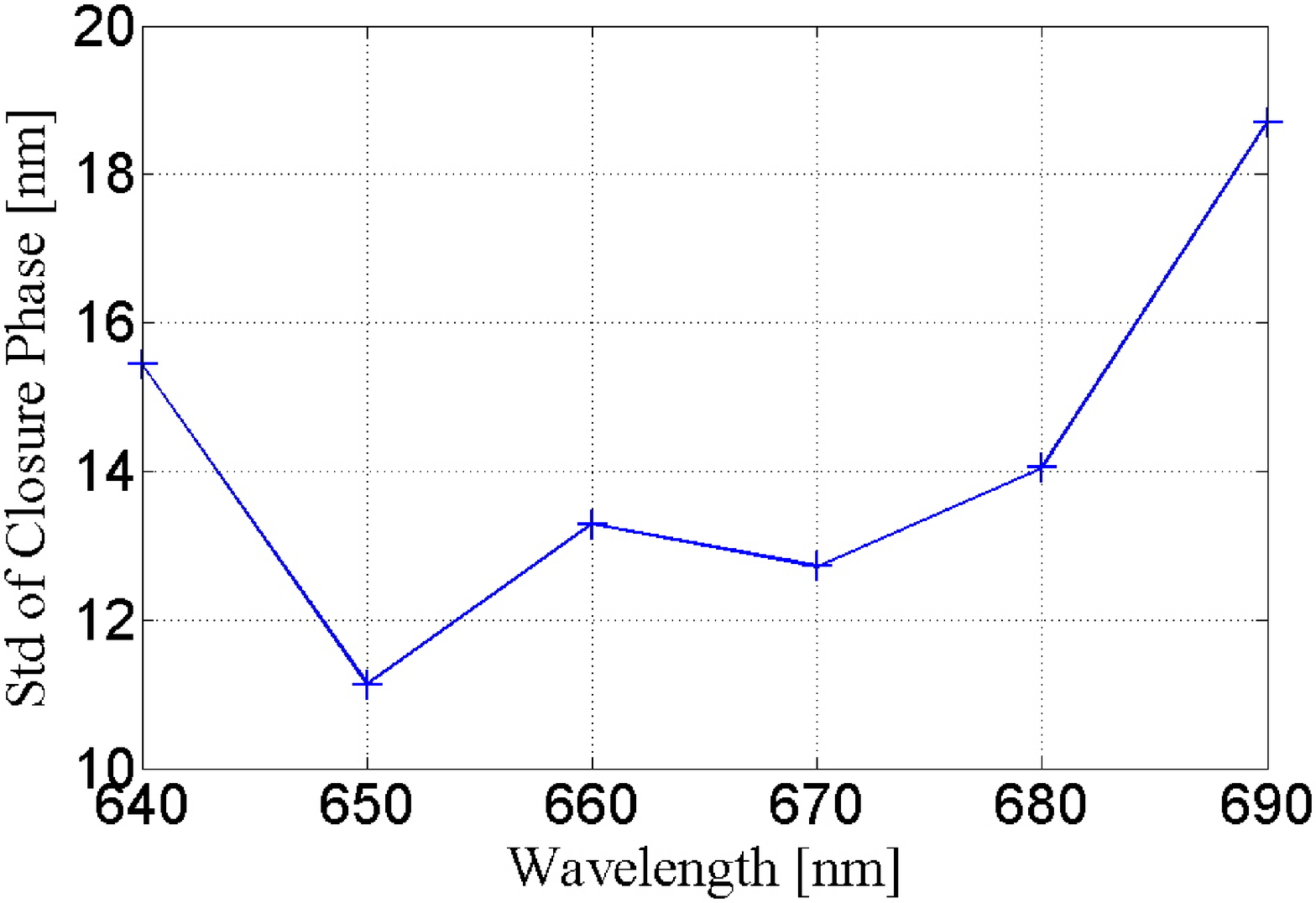}
   \end{tabular}
   \end{center}
   \caption[Integrated optic] 
   { \label{fig:visibilityclosure} 
Left: the measured peak visibilities as a function of wavelength for the three baselines of the experimental setup. Error bars are the standard deviation of 1000 measurements. Right: standard deviation of the closure phase as estimated from 1000 measurements with the DBC.}
   \end{figure}

We next analyzed the performance of the DBC in spectro-interferometry mode, by inserting 
the imaging spectrograph in front of the CCD camera.  
The light was dispersed along the horizontal direction (parallel to the lines of the waveguide array), so that 
we had a total free spectral range of 50 nm.
As a first test in the spectro-interferometry mode we verified that our system can deliver high visibilities throughout the spectral range. We set the optical path difference between the three
input beams to 0 for light at $\lambda_{\mathrm{c}}$= 660 nm. The optical path of the beams 1
and 3 was then modulated independently with a $\pm6\,\mu$m stroke with rates of 5.2
$\mu$m/s. The recorded the DBC output was used to retrieve a
time series of the normalized visibilities. The average values of the visibility and
its standard deviation are plot as a function of the wavelength in Fig. \ref{fig:visibilityclosure}(Left) for the
three baselines of our interferometric setup. The visibility of the three
baselines is constant within the error bars and its average value over all baselines
and all wavelengths is $0.88\pm0.06$, allowing a raw visibility dynamics of about 15.
Part of the residual variation of the visibilities over wavelength may be attributed
to the chromatic dispersion of the beam splitters used in the Mach-Zehnder interferometer.
With the same data we tested also the spectral uniformity of the DBC
method in retrieving optical path difference (OPD) data. The measured OPD
variation between beams 2 and 3 was 24.6 $\mu$m and was constant within 70 nm
across the investigated wavelength range. 

A crucial parameter for astronomical 
imaging is the stability of the closure phase. In our case the closure
phase is defined as $\Phi_{\mathrm{C}}=\Phi_{12}+\Phi_{23}-\Phi_{13}$, where $\Phi_{xy}$ is the optical path difference
between beams $x$ and $y$. 
We have measured the standard deviation 
of the closure phase for different wavelengths as illustrated in Fig.  \ref{fig:visibilityclosure}(Right). We
have obtained a stability better than 19 nm ($\lambda/36$) over the whole bandwidth of 50 nm,
with a minimum value of 11 nm ($\lambda/58$) at a wavelength of 650 nm. 
Notice that planar beam combiners for astronomical interferometry
were reported to have an absolute closure phase stability of 11 nm at a wavelength of 1550 nm 
($\lambda/144$) \cite{Benisty:2009}, a result in line with our observations.

\section{CONCLUSIONS AND PERSPECTIVES} 
In conclusion, we have shown how to operate a discrete beam combiner with polychromatic light. The chromatic variation of the coupling strength of the waveguide array is handled by dispersing the output discrete interference pattern with a compact imaging spectrometer. 
The setup can be used for low resolution spectro-interferometry. 

The DBC approach to astronomical interferometry has a perspective for the integrated, 'all-in-one' combination of large arrays of telescopes or sub-apertures, which cannot be delivered easily by the planar integrated optics. We mention that recent numerical searches found DBC configurations capable of combining up to 8 telescopes (Ronny Errmann, personal communication).  
Particularly attractive is the perspective to implement the DBC concept for mid-infrared wavelengths, where conventional integrated optics is at its infancy. By exploiting the technological platform of femtosecond laser writing, simple and robust beam combiners can be manufactured at lower cost per unit than the photolithographic approach.

Some technological developments are still necessary to improve the quality of the laser written waveguides, especially for what concerns their 
transparency. In fact, laser written waveguides have transmission losses in the visible in the order of 0.5 dB/cm and can reach 1-2 dB/cm in the mid-infrared\cite{Rodenas:2012}. Considering that the DBC sample length could be in the order of 5 to 10 cm, a considerable loss of energy is expected. In this context, post-processing treatment of the laser fabricated waveguides can result in a significant reduction of intrinsic losses \cite{Arriola:2013}.

For what concerns the writing of large arrays of waveguides, it should be noted that techniques for writing 'deep' into thick materials ($\sim 1$mm) would be needed. Tools for doing this (like active compensation of spatial and temporal aberrations of the writing pulse) are currently at reach of today's technology\cite{Stoian:2014}.

 


\bibliography{reportDBC}   

\begin{thebibliography}{10}

\bibitem{DdS:2007}
{Domiciano de Souza}, A., {Driebe}, T., {Chesneau}, O., {Hofmann}, K.-H.,
  {Kraus}, S., {Miroshnichenko}, A.~S., {Ohnaka}, K., {Petrov}, R.~G.,
  {Preisbisch}, T., {Stee}, P., {Weigelt}, G., {Lisi}, F., {Malbet}, F., and
  {Richichi}, A., ``{AMBER/VLTI and MIDI/VLTI spectro-interferometric
  observations of the B[e] supergiant CPD-57$^{\circ}$2874. Size and geometry
  of the circumstellar envelope in the near- and mid-IR},'' {\em A\&A}~{\bf
  464},  81--86 (Mar. 2007).

\bibitem{Lacour:2008}
{Lacour}, S., {Meimon}, S., {Thi{\'e}baut}, E., {Perrin}, G., {Verhoelst}, T.,
  {Pedretti}, E., {Schuller}, P.~A., {Mugnier}, L., {Monnier}, J., {Berger},
  J.~P., {Haubois}, X., {Poncelet}, A., {Le Besnerais}, G., {Eriksson}, K.,
  {Millan-Gabet}, R., {Ragland}, S., {Lacasse}, M., and {Traub}, W., ``{The
  limb-darkened Arcturus: imaging with the IOTA/IONIC interferometer},'' {\em
  A\&A}~{\bf 485},  561--570 (July 2008).

\bibitem{Benisty:2009}
Benisty, M., Berger, J.-P., Jocou, L., Labeye, P., Malbet, F., Perraut, K., and
  Kern, P., ``An integrated optics beam combiner for the second generation vlti
  instruments,'' {\em A\&A}~{\bf 498},  601--613 (2009).

\bibitem{Blind:2011}
{Blind}, N., {Boffin}, H.~M.~J., {Berger}, J.-P., {Le Bouquin}, J.-B.,
  {M{\'e}rand}, A., {Lazareff}, B., and {Zins}, G., ``{An incisive look at the
  symbiotic star SS Leporis. Milli-arcsecond imaging with PIONIER/VLTI},'' {\em
  A\&A}~{\bf 536},  A55 (Dec. 2011).

\bibitem{Rodenas:2012}
Rodenas, A., Martin, G., Arzeki, B., Psaila, N.~D., Jose, G., Jha, A., Labadie,
  L., Kern, P., Kar, A.~K., and Thomson, R.~R., ``Three-dimensional
  mid-infrared photonic circuits in chalcogenide glass,'' {\em Opt. Lett.}~{\bf
  37},  392--394 (2012).

\bibitem{Perrin:2006}
{Perrin}, G., {Lacour}, S., {Woillez}, J., and {Thi{\'e}baut}, {\'E}., ``{High
  dynamic range imaging by pupil single-mode filtering and remapping},'' {\em
  Month. Not. R. Astron. Soc.}~{\bf 373},  747--751 (Dec. 2006).

\bibitem{Minardi:2012b}
Minardi, S., Dreisow, F., Gr\"afe, M., Nolte, S., and Pertsch, T.,
  ``Three-dimensional photonic component for multichannel coherence
  measurements,'' {\em Opt. Lett.}~{\bf 37},  3030--3032 (2012).

\bibitem{Minardi:2010}
Minardi, S. and Pertsch, T., ``Interferometric beam combination with discrete
  optics,'' {\em Opt. Lett.}~{\bf 35},  3009--3011 (2010).

\bibitem{Saviauk:2013}
{Saviauk}, A., {Minardi}, S., {Dreisow}, F., {Nolte}, S., and {Pertsch}, T.,
  ``{3D-integrated optics component for astronomical spectro-interferometry},''
  {\em Appl. Opt.}~{\bf 52},  4556 (July 2013).

\bibitem{Christodoulides:2003}
{Christodoulides}, D.~N., {Lederer}, F., and {Silberberg}, Y., ``{Discretizing
  light behaviour in linear and nonlinear waveguide lattices},'' {\em
  Nature}~{\bf 424},  817--823 (Aug. 2003).

\bibitem{Tatulli:2007}
Tatulli, E., Millour, F., Chelli, A., Duvert, G., Acke, B., Utrera, O.~H.,
  Hofmann, K.-H., Kraus, S., Malbet, F., M\`ege, P., Petrov, R., Vannier, M.,
  Zins, G., Antonelli, P., Beckmann, U., Bresson, Y., Dugu\'e, M., Gennari, S.,
  Gl\"uck, L., Kern, P., Lagarde, S., Coarer, E.~L., Lisi, F., Perraut, K.,
  Puget, P., Rantakyr\"o, F., Robbe-Dubois, S., Roussel, A., Weigelt, G.,
  Accardo, M., Agabi, K., Altariba, E., Arezki, B., Aristidi, E., Baffa, C.,
  Behrend, J., Bl\"ocker, T., Bonhomme, S., Busoni, S., Cassaing, F., Clausse,
  J.-M., Colin, J., Connot, C., Delboulb\'e, A., de~Souza, A.~D., Driebe, T.,
  Feautrier, P., Ferruzzi, D., Forveille, T., Fossat, E., Foy, R.,
  Fraix-Burnet, D., Gallardo, A., Giani, E., Gil, C., Glentzlin, A., Heiden,
  M., Heininger, M., Kamm, D., Kiekebusch, M., Contel, D.~L., Contel, J.-M.~L.,
  Lesourd, T., Lopez, B., Lopez, M., Magnard, Y., Marconi, A., Mars, G.,
  Martinot-Lagarde, G., Mathias, P., Monin, J.-L., Mouillet, D., Mourard, D.,
  Nussbaum, E., Ohnaka, K., Pacheco, J., Perrier, C., Rabbia, Y., Rebattu, S.,
  Reynaud, F., Richichi, A., Robini, A., Sacchettini, M., Schertl, D.,
  Sch\"oller, M., Solscheid, W., Spang, A., Stee, P., Stefanini, P., Tallon,
  M., Tallon-Bosc, I., Tasso, D., Testi, L., Vakili, F., von~der L\"uhe, O.,
  Valtier, J.-C., and Ventura, N., ``Interferometric data reduction with
  amber/vlti. principle, estimators and illustration,'' {\em A\&A}~{\bf 464},
  29--42 (2007).

\bibitem{Minardi:2012}
Minardi, S., ``Photonic lattices for astronomical interferometry,'' {\em Month.
  Not. R. Astr. Soc.}~{\bf 422},  2656Ð2660 (2012).

\bibitem{Press}
{Press}, W.~H., {Teukolsky}, S.~A., {Vetterling}, W.~T., and {Flannery}, B.~P.,
   [{\em {Numerical recipes in C++ : the art of scientific
  computing}}{\nolinebreak\hspace{0.1em}]} (2002).

\bibitem{SnyderLove}
Snyder, A.~W. and Love, J.~D.,  [{\em Optical Waveguide
  Theory}{\nolinebreak\hspace{0.1em}]}, Kluwer, London (2007).

\bibitem{AllingtonSmith:2006}
{Allington-Smith}, J., ``{Basic principles of integral field spectroscopy},''
  {\em New Astron. Rev.}~{\bf 50},  244--251 (June 2006).

\bibitem{Bacon:2001}
{Bacon}, R., {Copin}, Y., {Monnet}, G., {Miller}, B.~W., {Allington-Smith},
  J.~R., {Bureau}, M., {Carollo}, C.~M., {Davies}, R.~L., {Emsellem}, E.,
  {Kuntschner}, H., {Peletier}, R.~F., {Verolme}, E.~K., and {de Zeeuw}, P.~T.,
  ``{The SAURON project - I. The panoramic integral-field spectrograph},'' {\em
  Month. Not. R. Astron. Soc.}~{\bf 326},  23--35 (Sept. 2001).

\bibitem{Felix}
{Bl{\"o}mer}, D., {Szameit}, A., {Dreisow}, F., {Schreiber}, T., {Nolte}, S.,
  and {T{\"u}nnermann}, A., ``{Nonlinear refractive index of fs-laser-written
  waveguides in fused silica},'' {\em Optics Express}~{\bf 14},  2151--2157
  (Mar. 2006).

\bibitem{Arriola:2013}
{Arriola}, A., {Gross}, S., {Jovanovic}, N., {Charles}, N., {Tuthill}, P.~G.,
  {Olaizola}, S.~M., {Fuerbach}, A., and {Withford}, M.~J., ``{Low bend loss
  waveguides enable compact, efficient 3D photonic chips},'' {\em Optics
  Express}~{\bf 21},  2978 (Feb. 2013).

\bibitem{Stoian:2014}
{Stoian}, R., {Colombier}, J.~P., {Mauclair}, C., {Cheng}, G., {Bhuyan}, M.~K.,
  {Velpula}, P.~K., and {Srisungsitthisunti}, P., ``{Spatial and temporal laser
  pulse design for material processing on ultrafast scales},'' {\em Applied
  Physics A: Materials Science \& Processing}~{\bf 114},  119--127 (Jan. 2014).

\end{thebibliography}
\bibliographystyle{spiebib}   

\end{document}